\documentclass[aps,twocolumn,amsmath,amssymb,prl,floatfix]{revtex4}

\usepackage{tabularx}
\usepackage{psfrag}
\usepackage{subfigure}
\usepackage{graphicx}
\usepackage{dcolumn}
\usepackage{bm}

\begin{document}

\title{Signatures of fractional statistics in noise experiments in quantum Hall fluids}

\author{Eun-Ah Kim}
\affiliation{Department of Physics, University of Illinois at
Urbana-Champaign, 1110 W. Green St., Urbana IL 61801-3080}
\author{Michael Lawler}
\affiliation{Department of Physics, University of Illinois at
Urbana-Champaign, 1110 W. Green St., Urbana IL 61801-3080}
\author{Smitha Vishveshwara}
\affiliation{Department of Physics, University of Illinois at
Urbana-Champaign, 1110 W. Green St., Urbana IL 61801-3080}
\author{Eduardo Fradkin}
\affiliation{Department of Physics, University of Illinois at
Urbana-Champaign, 1110 W. Green St., Urbana IL 61801-3080}

\date{\today}

\begin{abstract}
The elementary excitations of fractional quantum Hall (FQH) fluids are vortices with fractional statistics. Yet, this fundamental prediction has remained an open experimental challenge. Here we show that the cross current noise in a three-terminal tunneling experiment  of a two dimensional electron gas in the FQH regime can be used to detect directly the statistical angle of the excitations of these topological quantum fluids. We show that the noise also reveals signatures of exclusion statistics and of fractional charge. The vortices of Laughlin states should exhibit a ``bunching'' effect, while for higher states in the Jain sequences they should exhibit an ``anti-bunching'' effect.
\end{abstract}

\maketitle

The classification of fundamental particles in terms of their quantum statistics, Bose-Einstein (bosons) and Fermi-Dirac (fermions), is a fundamental law of Nature enshrined in the Principles of Quantum Mechanics. In the form of the {\em Spin-Statistics Theorem} it is also one of the basic {\em axioms} of Quantum Field Theory (and String Theory as well). 
However, it has long been known that other types of quantum statistics are also possible if the physical system has a reduced dimensionality.
Indeed, the possible existence of {\em anyons}\cite{leinaas77,wilczek82}, particles  with {\em fractional or braid } statistics, interpolating between bosons and fermions, is one of the most startling predictions of Quantum Mechanics in two space dimensions. 

The best experimental candidates for anyons presently known are the vortices (or quasiparticles (qp)) 
of a strongly interacting 2D electron gas (2DEG) in a strong magnetic field in the FQH regime\cite{tsui82,laughlin83,haldane83,halperin84,arovas84}.
In this regime, the 2DEG behaves as an incompressible 
dissipationless topological fluid exhibiting
the FQH effect \cite{tsui82,laughlin83}.  
The  quasiparticles of FQH fluids have remarkable 
properties\cite{laughlin83,haldane83,halperin84,arovas84}: 
 they are finite energy vortices (or solitons)
which carry  fractional charge and fractional (braid) statistics.
Although by now there is strong experimental evidence for fractional charge
\cite{ goldman95,picciotto97,saminadayar97,reznikov99,Yacoby04}, similar evidence is still lacking for fractional statistics (such experimental evidence has been reported very recently
~\cite{camino05}.)

There are two ways to think about statistics. A first way is through the concept of  {\em braiding (fractional) statistics}, in which the two particle wave function $\Psi({\bf r}_1, {\bf r}_2)$  acquires a statistical angle $\theta$ upon an adiabatic exchange process,
\begin{equation}
\Psi({\bf r}_1, {\bf r}_2)=e^{i\theta}\Psi({\bf r}_2, {\bf r}_1).
\label{eq:exchange}
\end{equation}
For fractional statistics to occur, the statistical angle $\theta$ should take values intermediate between that of bosons ($\theta\!=\!0$,  particles commute)  and fermions ($\theta\!=\!\pi$, particles anti-commute)\cite{wilczek82, arovas84}. 
The other way is to consider, in a finite size system,  the effect of the presence of particles to the subsequent addition of another particle. The Pauli Exclusion Principle
dictates that each fermion in a system reduces the available states to add extra fermions by $1$. In contrast, the presence of a boson does not affect the later addition of bosons at all. The natural consequence of  the presence of an anyon 
is an intermediate effect, {\it i.e.\/} a generalized exclusion principle\cite{haldane91,vanelburg98}.  
Unfortunately, neither braiding nor exclusion of anyons have  been unequivocally observed to date.

The difficulty in detecting fractional statistics lies in the extended nature of these 
quantum vortices. The adiabatic transport of one vortex around another is a conceptually nice {\em gedanken} experiment which is difficult to realize in practice. 
A number of interesting interferometers have been proposed to detect fractional statistics in FQH fluids \cite{jain93,chamon-freed-kivelson-sondhi-wen97,Kane03}, none of which has yet been realized experimentally, except possibly for the recent experiments of Ref.\cite{camino05}. Theoretical schemes to measure non-Abelian statistics have also been proposed.\cite{fradkin98,freedman05}

Here we propose an interferometer designed to detect {\em noise correlations} of the charge currents in a three-terminal experiment (a ``T- junction'') of a 2DEG in the FQH regime. We will show that the structure of the noise in the charge current cross-correlations in this T-junction provides a unique way to detect the fractional statistics of the vortices  directly
in realistic experimental conditions.  
This study was inspired  by Refs.~\cite{safi01,vishveshwara03} and we will compare our results with these references below. 

A zoo of different anyonic qp's have long been predicted to exist
for  each filling factor  $\nu$ (the number of electrons per flux quanta) exhibiting the FQH effect (See Table 1 for examples).  At each allowed $\nu$ the 2DEG displays a precise (fractional) quantization of the Hall conductance  $\sigma_{xy}=\nu e^2/h$. These quantum states are  distinct phases of the 2DEG distinguished by  a non-trivial internal or {\it topological order}, which is robust against arbitrary local perturbations \cite{wen-niu90,wen95}. The measurement of fractional statistics will directly prove the existence of topological order. In fully polarized 2DEGs the most prominent states lie in the Jain sequences \cite{jain89},
with filling fractions $\nu=p/(2np\!+\!1)$, including the Laughlin states\cite{laughlin83} (for $p\!=\!1$).
The statistical angle of vortices of Laughlin states lie in the range $\theta/\pi\!<\!1/2$, and are closer to that of bosons. In contrast, for the higher states in the Jain sequence, with $p\!>\!1$, the statistical angle of their vortices lie in the range 
$\theta/\pi\!>\!1/2$, which is closer to that of fermions \cite{lopez99, comment-wen} (See Table \ref{table:T1}). We show below that this difference leads to a ``bunching'' effect for Laughlin states and to an ``anti-bunching'' effect for non-Laughlin states.  In particular, in these sequences there are pairs of FQH states , {\it e.g.\/} $\nu=1/5$ and $\nu=2/5$, whose qp's have the same fractional charge but different fractional statistics. Hence, their qp's exhibit either bunching or anti-bunching behavior while having the same charge.

\begin{table}[h]
\newcolumntype{Y}{>{\centering\arraybackslash$}m{1.2cm}<{$}}
\newcolumntype{C}{>{\centering\arraybackslash$}m{2.3cm}<{$}}
\renewcommand{\arraystretch}{2}
\begin{tabular}{|Y||Y|Y|C|Y|}
\hline
  	&{\textrm{free}}\atop{\textrm{boson}} & p\!=\!1  & p\!>\!1& 
	{\textrm{free}}\atop{\textrm{fermion}}\\
\hline
 e^*/e& 0,2,\cdots & \nu & \displaystyle{\frac{\nu}{p}} & 1 \\
\hline
  \theta/\pi& 0 & \nu<\displaystyle\frac{1}{2} & \left(1\!-\!\displaystyle{\frac{2n}{p}}\nu\right)>\displaystyle\frac{1}{2} & 1\\
\hline
\end{tabular}
\label{table:qn}
\caption{Fractional charge $e^*$ and statistical angle $\theta$ for qp's of FQH states at filling factor $\nu = p/(2np\!+\!1)$ ($n$, $p$ are integers) (Ref.~\cite{lopez99}) compared with free bosons and free fermions. }
\label{table:T1}
\end{table} 
 
\begin{figure}[t]
\psfrag{0}{\small$0$}
\psfrag{1}{\small$1$}
\psfrag{2}{\small$2$}
\psfrag{V1}{\small$V_1$}
\psfrag{V2}{\small$V_2$}
\psfrag{V0}{\small$V_0$}
\psfrag{t1}{}
\psfrag{t2}{}
\psfrag{e+}{\footnotesize$S$}
\psfrag{e-}{\footnotesize$O$}
\includegraphics[width=0.44\textwidth]{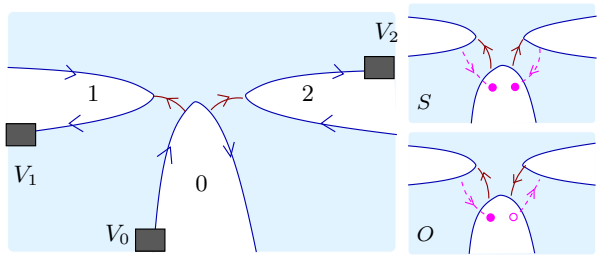}
\caption{The proposed 
T-junction\cite{comment}  and virtual processes for currents in the same ($S$)  and opposite ($O$) 
orientations. 
Blue lines represent edge states where
the edge state $0$ is held at potential $V$ relative to edges $1$ and $2$, {\it i.e.\/}
 $V_0-V_1=V_0-V_2=V$. Red solid lines show paths of two quasiparticle tunneling through a 
 FQH liquid from one edge state to two others and magenta dashed lines represent 
 virtual tunneling events.}
 \label{fig:hbt}
\end{figure}

Since the FQH liquids are incompressible,  it costs a finite amount of energy (the {\em gap}) to create a qp excitation in the 2D bulk of the 2DEG. 
However, these liquids support gapless one-dimensional chiral excitations at the edge
\cite{wen90,wen95},  opening a window for experimental probes.
In particular, tunneling between edges allows a practical access to single qp properties, provided the tunneling paths lie inside the FQH liquid (as the qp's can only exist within the FQH liquid). 
Single-point contact tunneling has been successfully used for the detection of fractional charge through shot 
noise measurements\cite{picciotto97,saminadayar97,reznikov99}.
However,  in order to stage an exchange between two qp's from multiple independent edges, one needs at least two point contacts. 
Further, if multiple edges are in point contacts with a single edge each adding/removing qp's 
to one edge, the generalized exclusion principle can manifest itself.  
A T-junction interferometer of the type shown in Fig.\ref{fig:hbt} is a minimal realization of such a situation.

In this T-junction interferometer, the qp's are driven to tunnel between one edge and two other edges. 
Top gates can be used to define and bring together three edges which 
are attached to external leads
by ohmic contacts.  
Upon setting the edge 
$0$ at relative voltage 
$V$ to the two others, qp's are driven to tunnel between these edge states.
The resulting tunneling current operator is 
\begin{equation}
I_j(t)\!=\!i\frac{e}{\hbar}^*\Gamma_j 
(e^{i\omega_0t}\psi_j^\dagger\psi_0 -e^{-i\omega_0t}\psi_0^\dagger\psi_j) , 
\label{eq:current}
\end{equation}
where $\Gamma_j$ is the tunneling amplitude between edges $0$ and $j\!=\!1,2$ 
respectively, and $\psi_0^\dagger$ and $\psi_j^\dagger$ 
are qp creation operators at the respective edges.
The Josephson frequency $\omega_0=e^*V/\hbar$  depends explicitly 
on the fractional charge. Each edge state, labelled by $l=0,1,2$, is a 1D chiral Luttinger liquid\cite{wen90,wen95},  whose equal-position qp propagator, at time $t$ and temperature $T$, is given by\cite{lopez99} 
\begin{equation}
\langle\psi_l(t)\psi^\dagger_l(0)\rangle =
\bigg|\frac{\pi\tau_0k_B T}{\sinh(\pi k_BTt)}\bigg|^{
K}e^{-i\frac{\theta}{2}{\textrm{sign}(t)}}. \label{eq:prop}
\end{equation}
Here $\psi^\dagger_l(t)$ is the qp creation operator for edge $l$, $\tau_0$ is a short time cutoff, and
$k_B$ is the Boltzmann constant.
Eq.\eqref{eq:prop} defines the scaling dimension of the qp 
operator ${\frac{K}{2}=\frac{1}{2p(2np+1)}}$. The fractional statistics, defined in Eq.~\eqref{eq:exchange}
for qp's of the same edge, is encoded in this propagator, through its explicit dependence on the statistical angle $\theta$.
For qp's of different edges, $l$ and $m$, 
the qp operators obey the (inter-edge) commutation rules  
\begin{equation}
\psi^\dagger_l\psi^\dagger_m\!=\!e^{-i\alpha_{lm}}\psi^\dagger_m\psi^\dagger_l,
\label{eq:klein}
\end{equation} 
with $\alpha_{02}\!=\!\alpha_{21}\!=\!\alpha_{01}\!=\!\theta$ and $\alpha_{lm}\!=
\!-\alpha_{ml}$.
Our goal is to find a measurable quantity which reflects the $\theta$ dependence in Eqs.\eqref{eq:prop} and \eqref{eq:klein}.

The normalized charge current cross-correlation $S(t)$ (noise) 
 ($\Delta I_l=I_l-\langle I_l \rangle$):
\begin{equation}
S(t)\equiv\frac{ \langle \Delta I_1(t)\Delta I_2(0)\rangle}
{\langle I_1\rangle\langle I_2\rangle} .
\label{eq:St}
\end{equation}
turns out to be the simplest quantity that can exhibit
the subtle signatures of fractional statistics\cite{safi01}, and distinguish them from
those of fractional charge. 
We have calculated this non-equilibrium noise, using the Schwinger-Keldysh
technique\cite{Chamon93,Chamon94} to lowest order in the tunneling amplitudes $\Gamma_j$.
The two virtual tunneling events in Fig.~\ref{fig:hbt}  are the 
lowest order non-vanishing contributions to the correlator $S(t)$ 
and they take place at all times allowed by causality. 
Depending on the relative orientations of the currents,
the virtual tunneling processes leave two distinct pairs of objects at edge $0$,
as shown in Fig.~\ref{fig:hbt}.
When the currents $I_1$ and $I_2$ have the same orientation (case $S$),  
a qp-qp pair (or a qh-qh pair) is left behind, while when
they have opposite orientations (case $O$), qp-qh pairs are left behind.
Comparing the statistical phase gain upon exchange of qp-qp pairs (or qh-qh pairs) in case $S$ and exchange of qp-qh pair in case $O$ prescribed in  Eqs.~\eqref{eq:prop} and \eqref{eq:klein}, 
we find that {\it all } the dependence in fractional statistics is captured by 
a factor of $\cos\theta$ coming from contributions of case $S$. 

The frequency spectrum of  $S(t)$ in Eq.~\eqref{eq:St} can be written as a sum of two 
functions, a ``direct'' and 
an ``exchange'' term, with a dependence on $\theta$ which is both simple and explicit:\cite{kim-prep}
\begin{equation}
\widetilde S\left(\frac{\omega}{\omega_0}\right)=A\left(\frac{\omega}{\omega_0}, \frac{\displaystyle T}{\displaystyle T_0}, K\right) + 
{\cos\theta}\ B\left(\frac{\omega}{\omega_0} , \frac{\displaystyle T}{\displaystyle T_0}, K\right),
\label{eq:ab}
\end{equation}
where $\omega$ is the frequency and $T_0\!=\!\hbar\omega_0/k_B$.\cite{comment} 
This expression displays a number of remarkable properties:
\renewcommand{\labelenumi}{\alph{enumi}.}
\begin{enumerate}
\item  
 This universal form, which applies for all Jain states,  is the lowest order 
 in perturbation theory in the tunneling amplitude. It is accurate provided the tunneling current is small (compared with the Hall current).
Non-universal terms contribute at higher orders.
\item Fractional charge and statistics play fundamentally distinct roles: fractional charge  enters Eq.~\eqref{eq:ab} through the Josephson frequency $\omega_0=e^*V/\hbar$ and fractional statistics through the $\cos\theta$ factor.
\item Given that  for  Laughlin states $\theta<\pi/2$ and
$\cos\theta>0$, the exchange term in Eq~(\ref{eq:ab}) provides largely positive (``bunching'' ) 
contributions to the noise whereas for non-Laughlin states with $\theta>\pi/2$, its contribution is largely negative (``anti-bunching'').  
\item That only case $S$ (Fig.~\ref{fig:hbt}) contributes to this factor can be viewed as a manifestation of a
generalized exclusion principle\cite{haldane91} since the virtual processes in this case involves adding a qp to edge $0$ in the presence of another.  
This is an intriguing observation
given that Eq.~\eqref{eq:ab} was derived using anyonic commutation rules prescribed by {\it braiding} statistics\cite{arovas84}.
\end{enumerate}

\begin{figure}[b]
\psfrag{S}{\footnotesize$\widetilde A_0$, $\widetilde B_0$ }
\psfrag{T}{\footnotesize$T/T_0$}
\psfrag{1}{\tiny$1$}
\psfrag{2}{\tiny$2$}
\psfrag{3}{\tiny$3$}
\psfrag{4}{\tiny$4$}
\psfrag{-4}{\tiny$-4$}
\psfrag{-8}{\tiny$-8$}
\includegraphics[width=0.36\textwidth]{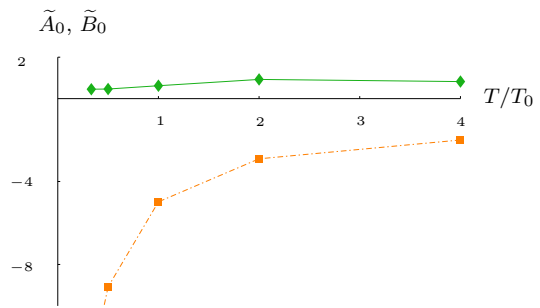}
\caption{\label{fig:Sw0} 
The temperature dependence the direct term $\widetilde A_0$ (orange dashed line) and the exchange term $\widetilde B_0$ (green solid line) defined in the text, for $\nu=2/5$  with $T_0\sim80{\text m}K$ ( $V\!=\!40\mu V$). This qualitative behavior is generic to all fractions. }
\end{figure}

We have studied (numerically) the behavior of the functions 
$\widetilde A(\omega/\omega_0, T/T_0,K)$  and $\widetilde B (\omega/\omega_0,T/T_0,K)$, whose details will be reported in a future publication\cite{kim-prep}. The salient results of this study, for $\omega\ll\omega_0$ and $\omega\sim\omega_0$, are as follows.
For  $\omega\ll\omega_0$, both  $\widetilde A$ and $\widetilde B$ show white noise-like  behavior, and hence we focus on zero frequency values $\widetilde A(\omega\!=\!0)\equiv\widetilde A_0$ and $\widetilde B(\omega\!=\!0)\equiv\widetilde B_0$ which depend only on the temperature (See Fig.~\ref{fig:Sw0}). 
$\widetilde A_0$ is negative due to the opposite orientation of the  currents in case $O$ of Fig.~\ref{fig:hbt} while $\widetilde B_0$ is positive as it only involves case $S$. Hence, $\widetilde S(\omega\!\!=\!\!0)\!=\!-|\widetilde A_0|\!+\!\cos\theta|\widetilde B_0|$. 
As shown in Fig.~\ref{fig:Sw0}, $-|\widetilde A_0|$ dominates the spectrum  at low temperatures ($T\ll T_0$) due to divergent contributions from case $O$ ($T^{-K}$ power law divergence). 
Nonetheless,  $\cos\theta|\widetilde B_0|$ significantly contributes to 
the total noise for higher temperatures of $2\lesssim T/T_0\lesssim4$. Thus, the statistical angle $\theta$ 
can be measured directly at a finite temperature.  

\begin{figure}[t]
 \subfigure[]
 {
\psfrag{S}{\footnotesize$\widetilde{S}(\omega/\omega_0)$}
\psfrag{S1by5}{\small$\nu\!=\!1/5$}
\psfrag{S2by5}{\small $\nu\!=\!2/5$}
\psfrag{w}{\footnotesize$\omega/\omega_0$}
\psfrag{20}{\footnotesize$20$}\psfrag{10}{\footnotesize$10$}\psfrag{-10}{\footnotesize$-10$}
\psfrag{-20}{\footnotesize$-20$}
\psfrag{0.5}{\footnotesize$0.5$}
\psfrag{1}{\footnotesize$1$}\psfrag{1.5}{\footnotesize$1.5$}
\includegraphics[width=0.35\textwidth]{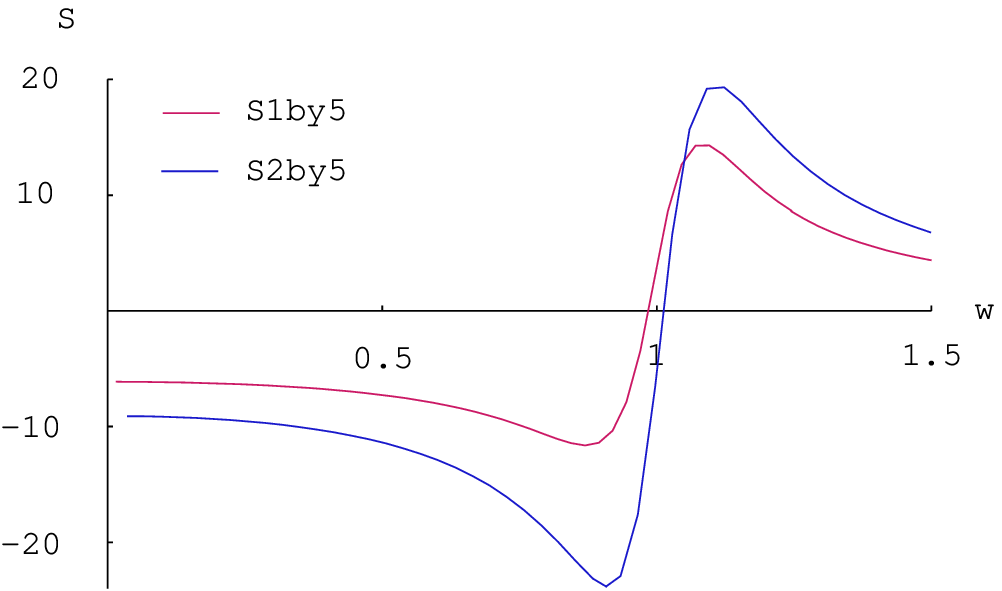}
}
\subfigure[]
{
\psfrag{e}{$e$}
\psfrag{e}{\footnotesize$\bar e$}
\psfrag{T}{\footnotesize$T/T_0$}
\psfrag{1by5}{\footnotesize$\bar e_{1/5}$}
\psfrag{2by5}{\footnotesize$\bar e_{2/5}$}
\psfrag{0.5}{\footnotesize$0.5$}
\psfrag{1}{\footnotesize$1$}
\psfrag{0.19}{\footnotesize$0.19$}
\psfrag{0.2}{\footnotesize$e^*(0.2)$}
\psfrag{0.21}{\footnotesize$0.21$}
\psfrag{0.22}{\footnotesize$0.22$}
\includegraphics[width=0.3\textwidth]{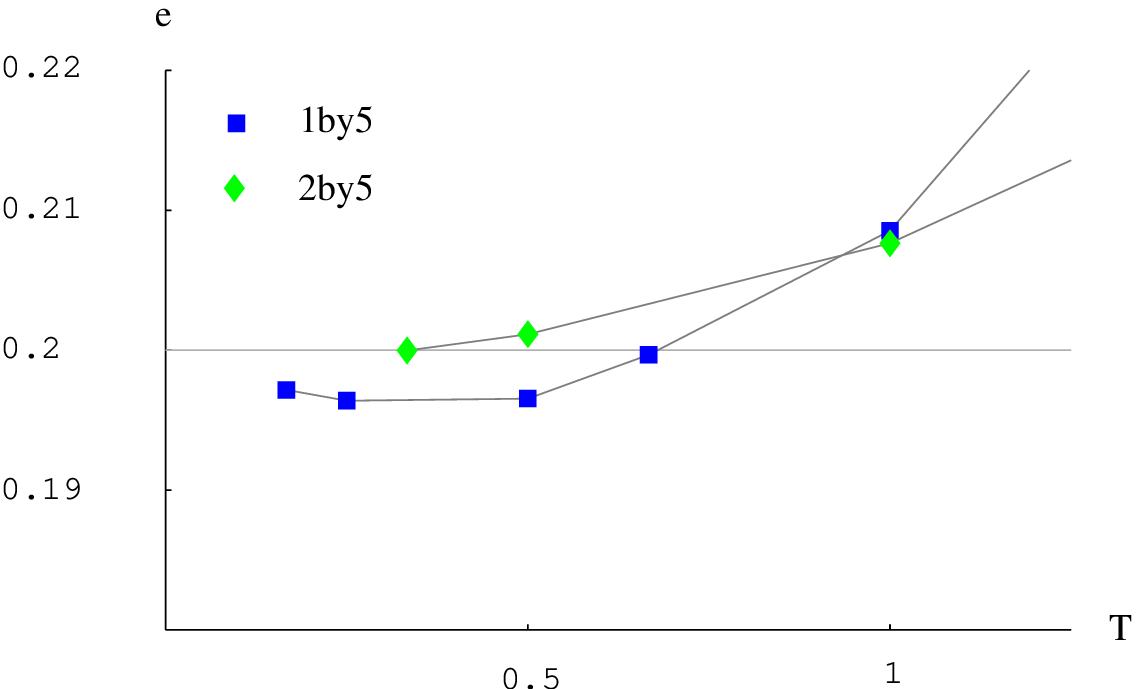}
}
\caption{\label{fig:Sw-peak} (a) $\widetilde{S}(\omega/\omega_0)$ for $\nu=1/5, 2/5$, for $T=20 \text{m}K, 40 \text{m}K$ respectively with $V\!\sim\!40\mu V$. 
$\widetilde{S}(\omega/\omega_0=1)$ is positive for the Laughlin state $\nu=1/5$ and negative for the Jain state $\nu=2/5$. (b) The effective charge $\bar e (T)$ in units of $e$ is determined by the crossing points of (a) for $\nu\!=\!1/5$ and $\nu\!=\!2/5$.
}
\end{figure}

Near the Josephson frequency $\omega\!\sim\!\omega_0$,  $\widetilde A$ and $\widetilde B$ have qualitatively different sharp features at low temperatures ($T\ll T_0$): $\widetilde A(\omega/\omega_0)$ changes sign and crosses the frequency axis steeply, while
$\widetilde B(\omega/\omega_0)$ develops a peak  of width $2K k_BT/\hbar\omega_0$.
These sharp features provide a direct way to measure the fractional charge\cite{Chamon93,Chamon94}, through  the Josephson frequency. They can be used to distinguish between two filling factors with the same fractional charge but different statistics, {\it e.g.\/} $\nu=1/5$ and $\nu=2/5$.  Since the direct term nearly vanishes close to the Josephson frequency, the exchange term plays a significant role in the total correlation $\widetilde S(\omega/\omega_0\!=\!1)$ by shifting it to a positive (negative) value for Laughlin (non-Laughlin) states as shown in Fig.~\ref{fig:Sw-peak}(a). The information on statistics and charge contained in the noise  near the Josephson frequency is summarized in Fig.~\ref{fig:Sw-peak}(b),  where we
introduced the  ``effective charge'' $\bar e \equiv\hbar\bar\omega/V$ obtained from the crossing point of  $\widetilde S$  {\em i.e. } $\widetilde S(\bar\omega/\omega_0)=0$. 
Comparing  $\bar e$'s for $\nu\!=\!1/5$ and for $\nu\!=\!2/5$ we find them satisfying $\bar e_{1/5}\!<\!e^*\!<\!\bar e_{2/5}$ over a broad range of (low) temperatures, however both approaching the fractional charge $e^*\!=\!e/5$ as $T\to 0$.
 This different asymptotic behavior 
is a  consequence of their different statistical angles: $\theta\!=\!\pi/5$ (bunching) for $\nu=1/5$ and $\theta\!=\!3\pi/5$ (anti-bunching) for $\nu\!=\!2/5$.
This (weak) effect also shows that the vortices of these FQH states have different fractional statistics in spite of having the same fractional charge.

In summary, we have shown that the cross-current  noise in a T-junction can be used to directly measure  the statistical angle $\theta$ independently from the fractional charge, and conceptually, from the filling factor $\nu$.
The low temperature and low frequency results show that there is a  quantitatively significant statistics-dependence of this noise. We have shown how to measure both fractional charge and statistics near  the Josephson frequency. Our results are independent of non-universal short time physics. \textcite{safi01} considered a similar setup  for Laughlin states at $T=0$, and  obtained an expression for $\widetilde S(\omega\!=\!0)$ for case $O$ alone as a function of $\nu$ (for Laughlin states one cannot distinguish $\nu$ from $K$ or $e^*/e$ or $\theta/\pi$).  At {\em finite temperatures}, case $S$ (whose contribution is negligible next to case $O$ at zero temperature) brings in an explicit statistics dependence which can be distinguished from the effect of the fractional charge and the scaling dimension for non-Laughlin states.  A closely related setup is the four terminal case studied (only for Laughlin states at $T=0$) by Vishveshwara [\onlinecite{vishveshwara03}]. Even in this case, an actual measurement of the cross-correlation functions will involve the quantity $S(t)$ we calculated here. 

The current state-of-the-art noise experiments \cite{reznikov99,chung03}
operate at bias voltages of $10-150 \mu V$ (corresponding to a Josephson 
frequency $\omega_0 \sim 1$ GHz), and temperatures of $10-100mK$. 
These parameters access the ratios of $T/T_0$ in the range shown here, 
making our proposal quite feasible. To remain in the perturbative regime 
the voltage $V$ has to be large compared to a crossover scale determined 
by the tunneling amplitudes, so that the current is a small fraction 
($\sim 0.1$ ) of the Hall current at the same voltage, but still with a 
measurable noise.
Some recent experiments 
are consistent with fractional 
statistics. A somewhat large qp charge, extracted 
in recent shot noise
experiments, suggests bunching behavior.\cite{chung03} 
Aharonov-Bohm oscillations in an anti-dot with a 
super-period have
been interpreted as evidence for fractional statistics.\cite{camino05} 

\noindent
{\bf Acknowledgments}:
This work was supported in part by the National
Science Foundation through the grants DMR 0442537 (EK, ML, EF), DOE-MRL DEFG02-91-ER45439 (SV). We thank F.D.M. Haldane, M. Heiblum, and especially C. Chamon for important comments.


\begin{thebibliography}{33}
\expandafter\ifx\csname natexlab\endcsname\relax\def\natexlab#1{#1}\fi
\expandafter\ifx\csname bibnamefont\endcsname\relax
  \def\bibnamefont#1{#1}\fi
\expandafter\ifx\csname bibfnamefont\endcsname\relax
  \def\bibfnamefont#1{#1}\fi
\expandafter\ifx\csname citenamefont\endcsname\relax
  \def\citenamefont#1{#1}\fi
\expandafter\ifx\csname url\endcsname\relax
  \def\url#1{\texttt{#1}}\fi
\expandafter\ifx\csname urlprefix\endcsname\relax\def\urlprefix{URL }\fi
\providecommand{\bibinfo}[2]{#2}
\providecommand{\eprint}[2][]{\url{#2}}

\bibitem[{\citenamefont{Leinaas and Myerheim}(1977)}]{leinaas77}
\bibinfo{author}{\bibfnamefont{J.}~\bibnamefont{Leinaas}} \bibnamefont{and}
  \bibinfo{author}{\bibfnamefont{J.}~\bibnamefont{Myerheim}},
  \bibinfo{journal}{Nuovo Cimento Soc. Ital. Fis.}
  \textbf{\bibinfo{volume}{37B}}, \bibinfo{pages}{1} (\bibinfo{year}{1977}).

\bibitem[{\citenamefont{Wilczek}(1982)}]{wilczek82}
\bibinfo{author}{\bibfnamefont{F.}~\bibnamefont{Wilczek}},
  \bibinfo{journal}{Phys. Rev. Lett.} \textbf{\bibinfo{volume}{48}},
  \bibinfo{pages}{1144} (\bibinfo{year}{1982}).

\bibitem[{\citenamefont{Tsui et~al.}(1982)\citenamefont{Tsui, St\"ormer, and
  Gossard}}]{tsui82}
\bibinfo{author}{\bibfnamefont{D.~C.} \bibnamefont{Tsui}},
  \bibinfo{author}{\bibfnamefont{H.~L.} \bibnamefont{St\"ormer}},
  \bibnamefont{and} \bibinfo{author}{\bibfnamefont{A.~C.}
  \bibnamefont{Gossard}}, \bibinfo{journal}{Phys. Rev. Lett.}
  \textbf{\bibinfo{volume}{48}}, \bibinfo{pages}{1559} (\bibinfo{year}{1982}).

\bibitem[{\citenamefont{Laughlin}(1983)}]{laughlin83}
\bibinfo{author}{\bibfnamefont{R.~B.} \bibnamefont{Laughlin}},
  \bibinfo{journal}{Phys. Rev. Lett.} \textbf{\bibinfo{volume}{50}},
  \bibinfo{pages}{1395} (\bibinfo{year}{1983}).

\bibitem[{\citenamefont{Haldane}(1983)}]{haldane83}
\bibinfo{author}{\bibfnamefont{F.~D.~M.} \bibnamefont{Haldane}},
  \bibinfo{journal}{Phys. Rev. Lett.} \textbf{\bibinfo{volume}{51}},
  \bibinfo{pages}{605} (\bibinfo{year}{1983}).

\bibitem[{\citenamefont{Halperin}(1984)}]{halperin84}
\bibinfo{author}{\bibfnamefont{B.~I.} \bibnamefont{Halperin}},
  \bibinfo{journal}{Phys. Rev. Lett.} \textbf{\bibinfo{volume}{52}},
  \bibinfo{pages}{1583} (\bibinfo{year}{1984}).

\bibitem[{\citenamefont{Arovas et~al.}(1984)\citenamefont{Arovas, Schrieffer,
  and Wilczek}}]{arovas84}
\bibinfo{author}{\bibfnamefont{D.}~\bibnamefont{Arovas}},
  \bibinfo{author}{\bibfnamefont{J.~R.} \bibnamefont{Schrieffer}},
  \bibnamefont{and} \bibinfo{author}{\bibfnamefont{F.}~\bibnamefont{Wilczek}},
  \bibinfo{journal}{Phys. Rev. Lett.} \textbf{\bibinfo{volume}{53}},
  \bibinfo{pages}{722} (\bibinfo{year}{1984}).

\bibitem[{\citenamefont{Goldman and Su}(1995)}]{goldman95}
\bibinfo{author}{\bibfnamefont{V.~J.} \bibnamefont{Goldman}} \bibnamefont{and}
  \bibinfo{author}{\bibfnamefont{B.}~\bibnamefont{Su}},
  \bibinfo{journal}{Science} \textbf{\bibinfo{volume}{267}},
  \bibinfo{pages}{1010} (\bibinfo{year}{1995}).

\bibitem[{\citenamefont{de~Picciotto et~al.}(1997)\citenamefont{de~Picciotto,
  Reznikov, Heiblum, Umansky, Bunin, and Mahalu}}]{picciotto97}
\bibinfo{author}{\bibfnamefont{R.}~\bibnamefont{de~Picciotto}},
  \bibinfo{author}{\bibfnamefont{M.}~\bibnamefont{Reznikov}},
  \bibinfo{author}{\bibfnamefont{M.}~\bibnamefont{Heiblum}},
  \bibinfo{author}{\bibfnamefont{V.}~\bibnamefont{Umansky}},
  \bibinfo{author}{\bibfnamefont{G.}~\bibnamefont{Bunin}}, \bibnamefont{and}
  \bibinfo{author}{\bibfnamefont{D.}~\bibnamefont{Mahalu}},
  \bibinfo{journal}{Nature} \textbf{\bibinfo{volume}{389}},
  \bibinfo{pages}{162} (\bibinfo{year}{1997}).

\bibitem[{\citenamefont{Saminadayar et~al.}(1997)\citenamefont{Saminadayar,
  Glattli, Jin, and Etienne}}]{saminadayar97}
\bibinfo{author}{\bibfnamefont{L.}~\bibnamefont{Saminadayar}},
  \bibinfo{author}{\bibfnamefont{D.~C.} \bibnamefont{Glattli}},
  \bibinfo{author}{\bibfnamefont{Y.}~\bibnamefont{Jin}}, \bibnamefont{and}
  \bibinfo{author}{\bibfnamefont{B.}~\bibnamefont{Etienne}},
  \bibinfo{journal}{Phys. Rev. Lett.} \textbf{\bibinfo{volume}{79}},
  \bibinfo{pages}{2526} (\bibinfo{year}{1997}).

\bibitem[{\citenamefont{Reznikov et~al.}(1999)\citenamefont{Reznikov,
  de~Picciotto, Griffiths, Heiblum, and Umansky}}]{reznikov99}
\bibinfo{author}{\bibfnamefont{M.}~\bibnamefont{Reznikov}},
  \bibinfo{author}{\bibfnamefont{R.}~\bibnamefont{de~Picciotto}},
  \bibinfo{author}{\bibfnamefont{T.~G.} \bibnamefont{Griffiths}},
  \bibinfo{author}{\bibfnamefont{M.}~\bibnamefont{Heiblum}}, \bibnamefont{and}
  \bibinfo{author}{\bibfnamefont{V.}~\bibnamefont{Umansky}},
  \bibinfo{journal}{Nature} \textbf{\bibinfo{volume}{399}},
  \bibinfo{pages}{238} (\bibinfo{year}{1999}).

\bibitem[{\citenamefont{Martin et~al.}(2004)\citenamefont{Martin, Ilani,
  Verdene, Smet, Umansky, Mahalu, Schuh, Abstreiter, and Yacoby}}]{Yacoby04}
\bibinfo{author}{\bibfnamefont{J.}~\bibnamefont{Martin}},
  \bibinfo{author}{\bibfnamefont{S.}~\bibnamefont{Ilani}},
  \bibinfo{author}{\bibfnamefont{B.}~\bibnamefont{Verdene}},
  \bibinfo{author}{\bibfnamefont{J.}~\bibnamefont{Smet}},
  \bibinfo{author}{\bibfnamefont{V.}~\bibnamefont{Umansky}},
  \bibinfo{author}{\bibfnamefont{D.}~\bibnamefont{Mahalu}},
  \bibinfo{author}{\bibfnamefont{D.}~\bibnamefont{Schuh}},
  \bibinfo{author}{\bibfnamefont{G.}~\bibnamefont{Abstreiter}},
  \bibnamefont{and} \bibinfo{author}{\bibfnamefont{A.}~\bibnamefont{Yacoby}},
  \bibinfo{journal}{Science} \textbf{\bibinfo{volume}{305}},
  \bibinfo{pages}{980} (\bibinfo{year}{2004}).

\bibitem[{\citenamefont{Camino et~al.}()\citenamefont{Camino, Zhou, and
  Goldman}}]{camino05}
\bibinfo{author}{\bibfnamefont{F.~E.} \bibnamefont{Camino}},
  \bibinfo{author}{\bibfnamefont{W.}~\bibnamefont{Zhou}}, \bibnamefont{and}
  \bibinfo{author}{\bibfnamefont{V.~J.} \bibnamefont{Goldman}},
  \bibinfo{note}{unpublished, cond-mat/052406}.

\bibitem[{\citenamefont{Haldane}(1991)}]{haldane91}
\bibinfo{author}{\bibfnamefont{F.~D.~M.} \bibnamefont{Haldane}},
  \bibinfo{journal}{Phys. Rev. Lett.} \textbf{\bibinfo{volume}{67}},
  \bibinfo{pages}{937} (\bibinfo{year}{1991}).

\bibitem[{\citenamefont{van Elburg and Schoutens}(1998)}]{vanelburg98}
\bibinfo{author}{\bibfnamefont{R.~A.~J.} \bibnamefont{van Elburg}}
  \bibnamefont{and}
  \bibinfo{author}{\bibfnamefont{K.}~\bibnamefont{Schoutens}},
  \bibinfo{journal}{Phys. Rev. B} \textbf{\bibinfo{volume}{58}},
  \bibinfo{pages}{15704} (\bibinfo{year}{1998}).

\bibitem[{\citenamefont{Jain et~al.}(1993)\citenamefont{Jain, Kivelson, and
  Thouless}}]{jain93}
\bibinfo{author}{\bibfnamefont{J.~K.} \bibnamefont{Jain}},
  \bibinfo{author}{\bibfnamefont{S.~A.} \bibnamefont{Kivelson}},
  \bibnamefont{and} \bibinfo{author}{\bibfnamefont{D.~J.}
  \bibnamefont{Thouless}}, \bibinfo{journal}{Phys. Rev. Lett.}
  \textbf{\bibinfo{volume}{71}}, \bibinfo{pages}{3004} (\bibinfo{year}{1993}).

\bibitem[{\citenamefont{de~C.~Chamon et~al.}(1997)\citenamefont{de~C.~Chamon,
  Freed, Kivelson, Sondhi, and Wen}}]{chamon-freed-kivelson-sondhi-wen97}
\bibinfo{author}{\bibfnamefont{C.}~\bibnamefont{de~C.~Chamon}},
  \bibinfo{author}{\bibfnamefont{D.~E.} \bibnamefont{Freed}},
  \bibinfo{author}{\bibfnamefont{S.~A.} \bibnamefont{Kivelson}},
  \bibinfo{author}{\bibfnamefont{S.~L.} \bibnamefont{Sondhi}},
  \bibnamefont{and} \bibinfo{author}{\bibfnamefont{X.~G.} \bibnamefont{Wen}},
  \bibinfo{journal}{Phys. Rev. B} \textbf{\bibinfo{volume}{55}},
  \bibinfo{pages}{2331} (\bibinfo{year}{1997}).

\bibitem[{\citenamefont{Kane}(2003)}]{Kane03}
\bibinfo{author}{\bibfnamefont{C.~L.} \bibnamefont{Kane}},
  \bibinfo{journal}{Phys. Rev. Lett.} \textbf{\bibinfo{volume}{90}},
  \bibinfo{pages}{226802} (\bibinfo{year}{2003}).

\bibitem[{\citenamefont{Fradkin et~al.}(1998)\citenamefont{Fradkin, Nayak,
  Tsvelik, and Wilczek}}]{fradkin98}
\bibinfo{author}{\bibfnamefont{E.}~\bibnamefont{Fradkin}},
  \bibinfo{author}{\bibfnamefont{C.}~\bibnamefont{Nayak}},
  \bibinfo{author}{\bibfnamefont{A.}~\bibnamefont{Tsvelik}}, \bibnamefont{and}
  \bibinfo{author}{\bibfnamefont{F.}~\bibnamefont{Wilczek}},
  \bibinfo{journal}{Nucl. Phys. B} \textbf{\bibinfo{volume}{516}},
  \bibinfo{pages}{704} (\bibinfo{year}{1998}).

\bibitem[{\citenamefont{Freedman et~al.}(2005)\citenamefont{Freedman, Nayak,
  and {Das Sarma}}}]{freedman05}
\bibinfo{author}{\bibfnamefont{M.}~\bibnamefont{Freedman}},
  \bibinfo{author}{\bibfnamefont{C.}~\bibnamefont{Nayak}}, \bibnamefont{and}
  \bibinfo{author}{\bibfnamefont{S.}~\bibnamefont{{Das Sarma}}},
  \bibinfo{journal}{Phys. Rev. Lett.} \textbf{\bibinfo{volume}{94}},
  \bibinfo{pages}{1} (\bibinfo{year}{2005}).

\bibitem[{\citenamefont{Safi et~al.}(2001)\citenamefont{Safi, Devillard, and
  Martin}}]{safi01}
\bibinfo{author}{\bibfnamefont{I.}~\bibnamefont{Safi}},
  \bibinfo{author}{\bibfnamefont{P.}~\bibnamefont{Devillard}},
  \bibnamefont{and} \bibinfo{author}{\bibfnamefont{T.}~\bibnamefont{Martin}},
  \bibinfo{journal}{Phys. Rev. Lett.} \textbf{\bibinfo{volume}{86}},
  \bibinfo{pages}{4628} (\bibinfo{year}{2001}).

\bibitem[{\citenamefont{Vishveshwara}(2003)}]{vishveshwara03}
\bibinfo{author}{\bibfnamefont{S.}~\bibnamefont{Vishveshwara}},
  \bibinfo{journal}{Phys. Rev. Lett.} \textbf{\bibinfo{volume}{91}},
  \bibinfo{pages}{196803} (\bibinfo{year}{2003}).

\bibitem[{\citenamefont{Wen and Niu}(1990)}]{wen-niu90}
\bibinfo{author}{\bibfnamefont{X.~G.} \bibnamefont{Wen}} \bibnamefont{and}
  \bibinfo{author}{\bibfnamefont{Q.}~\bibnamefont{Niu}},
  \bibinfo{journal}{Phys. Rev. B} \textbf{\bibinfo{volume}{41}},
  \bibinfo{pages}{9377} (\bibinfo{year}{1990}).

\bibitem[{\citenamefont{Wen}(1995)}]{wen95}
\bibinfo{author}{\bibfnamefont{X.~G.} \bibnamefont{Wen}},
  \bibinfo{journal}{Advances in Physics} \textbf{\bibinfo{volume}{44}},
  \bibinfo{pages}{405} (\bibinfo{year}{1995}).

\bibitem[{\citenamefont{Jain}(1989)}]{jain89}
\bibinfo{author}{\bibfnamefont{J.~K.} \bibnamefont{Jain}},
  \bibinfo{journal}{Phys. Rev. Lett.} \textbf{\bibinfo{volume}{63}},
  \bibinfo{pages}{199} (\bibinfo{year}{1989}).

\bibitem[{\citenamefont{L\'opez and Fradkin}(1999)}]{lopez99}
\bibinfo{author}{\bibfnamefont{A.}~\bibnamefont{L\'opez}} \bibnamefont{and}
  \bibinfo{author}{\bibfnamefont{E.}~\bibnamefont{Fradkin}},
  \bibinfo{journal}{Phys. Rev. B} \textbf{\bibinfo{volume}{59}},
  \bibinfo{pages}{15323} (\bibinfo{year}{1999}).

\bibitem[{com({\natexlab{a}})}]{comment-wen}
\bibinfo{note}{Alternative descriptions ({\it e.g.\/} Ref. \cite{wen95})
  predict somewhat different spectrum of qp's and the proposed experiment can
  be used to distinguish different predictions.}

\bibitem[{\citenamefont{Wen}(1990)}]{wen90}
\bibinfo{author}{\bibfnamefont{X.~G.} \bibnamefont{Wen}},
  \bibinfo{journal}{Phys. Rev. B} \textbf{\bibinfo{volume}{41}},
  \bibinfo{pages}{12838} (\bibinfo{year}{1990}).

\bibitem[{com({\natexlab{b}})}]{comment}
\bibinfo{note}{We assume that, at edge $0$, the two qp's (or the qp-qh pair)
  are created (destroyed) at a distance $d$ small compared to the distance to
  edges $1$ and $2$. For $d>0$, Eq.\eqref{eq:ab} also depends on $e^* V d/\hbar
  v_F$. here we have set $d=0$.}

\bibitem[{\citenamefont{{de C. Chamon} et~al.}(1993)\citenamefont{{de C.
  Chamon}, Freed, and Wen}}]{Chamon93}
\bibinfo{author}{\bibfnamefont{C.}~\bibnamefont{{de C. Chamon}}},
  \bibinfo{author}{\bibfnamefont{D.~E.} \bibnamefont{Freed}}, \bibnamefont{and}
  \bibinfo{author}{\bibfnamefont{X.~G.} \bibnamefont{Wen}},
  \bibinfo{journal}{Phys. Rev. B} \textbf{\bibinfo{volume}{51}},
  \bibinfo{pages}{2363} (\bibinfo{year}{1993}).

\bibitem[{\citenamefont{{de C. Chamon} et~al.}(1994)\citenamefont{{de C.
  Chamon}, Freed, and Wen}}]{Chamon94}
\bibinfo{author}{\bibfnamefont{C.}~\bibnamefont{{de C. Chamon}}},
  \bibinfo{author}{\bibfnamefont{D.~E.} \bibnamefont{Freed}}, \bibnamefont{and}
  \bibinfo{author}{\bibfnamefont{X.~G.} \bibnamefont{Wen}},
  \bibinfo{journal}{Phys. Rev. B} \textbf{\bibinfo{volume}{53}},
  \bibinfo{pages}{4033} (\bibinfo{year}{1994}).

\bibitem[{\citenamefont{Kim et~al.}(2005)\citenamefont{Kim, Lawler,
  Vishveshwara, and Fradkin}}]{kim-prep}
\bibinfo{author}{\bibfnamefont{E.-A.} \bibnamefont{Kim}},
  \bibinfo{author}{\bibfnamefont{M.}~\bibnamefont{Lawler}},
  \bibinfo{author}{\bibfnamefont{S.}~\bibnamefont{Vishveshwara}},
  \bibnamefont{and} \bibinfo{author}{\bibfnamefont{E.}~\bibnamefont{Fradkin}}
  (\bibinfo{year}{2005}), \bibinfo{note}{in preparation}.

\bibitem[{\citenamefont{Chung et~al.}(2003)\citenamefont{Chung, Heiblum, and
  Umansky}}]{chung03}
\bibinfo{author}{\bibfnamefont{Y.~C.} \bibnamefont{Chung}},
  \bibinfo{author}{\bibfnamefont{M.}~\bibnamefont{Heiblum}}, \bibnamefont{and}
  \bibinfo{author}{\bibfnamefont{V.}~\bibnamefont{Umansky}},
  \bibinfo{journal}{Phys. Rev. Lett.} \textbf{\bibinfo{volume}{91}},
  \bibinfo{pages}{216804} (\bibinfo{year}{2003}).

\end{thebibliography}
\end{document}